%% file: eprint.tex
\documentclass[12pt]{article}
\usepackage{graphicx}
\usepackage{lineno}
\usepackage{url}
\usepackage{hyperref}
\newcommand\pubnumber{CMS CR-2021/299}
\newcommand\pubdate{\today}

\def\institute{Deutsches Elektronen-Synchrotron (DESY), Hamburg, Germany}

\def\Title#1{\begin{center} {\Large #1 } \end{center}}
\def\Author#1{\begin{center}{ \sc #1} \end{center}}
\def\Address#1{\begin{center}{ \it #1} \end{center}}

\newcommand\pubblock{\rightline{\begin{tabular}{l} \pubnumber\\
         \pubdate  \end{tabular}}}
\newenvironment{Abstract}{\begin{quotation}  }{\end{quotation}}
\newenvironment{Presented}{\begin{quotation} \begin{center} 
             PRESENTED AT\end{center}\bigskip 
      \begin{center}\begin{large}}{\end{large}\end{center} \end{quotation}}

\input econfmacros.tex
\newcommand{\ttbar}{\ensuremath{\mathrm{t\bar{t}}}}
\newcommand{\fbinv}{\ensuremath{\mathrm{fb^{-1}}}}
\newcommand{\yukawa}{\ensuremath{\mathrm{Y_{t}}}}
\newcommand{\mttbar}{\ensuremath{m_\mathrm{{t\bar{t}}}}}
\newcommand{\mtmc}{\ensuremath{m_\mathrm{{t}}^\mathrm{{MC}}}}

\begin{document}
\begin{titlepage}
\pubblock

\vfill
\Title{Standard model parameters from top quark measurements\\ at LHC with ATLAS and CMS}
\vfill
\Author{Sebastian Wuchterl\\on behalf of the ATLAS and CMS Collaborations\footnote{Copyright [2021] CERN for the benefit of the ATLAS and CMS Collaborations. Reproduction of this article or parts of it is allowed as specified in the CC-BY-4.0 license}}
\Address{\institute}
\vfill
\begin{Abstract}
Theoretical predictions for standard model (SM) processes involving top quarks, such as top quark-antiquark pair or single top production, depend on fundamental SM parameters like the strong coupling constant or the top quark mass. By confronting predictions with measurements performed at the ATLAS and CMS Collaboration using data collected at the CERN LHC in the second data taking period, these parameters can be determined precisely. In these proceedings, recent results measuring SM and top quark properties are presented.

\end{Abstract}
\vfill
\begin{Presented}
$14^\mathrm{th}$ International Workshop on Top Quark Physics\\
(videoconference), 13--17 September, 2021
\end{Presented}
\vfill
\end{titlepage}
\def\thefootnote{\fnsymbol{footnote}}
\setcounter{footnote}{0}

\section{Introduction}

At the CERN LHC, top quarks are predominantly produced in top quark-antiquark pair (\ttbar) production via the gluon fusion mechanism. But with the increased data set available from the second data taking period (LHC Run II) (2015-2018) at a center-of-mass-energy of 13\,TeV, also processes like single top \textit{t}-channel production can be studied in experiments with high precision. Properties of the top quark, such as its large mass or its Yukawa coupling close to unity to the Higgs boson, imply that the top quark plays a special role within the standard model (SM), especially in the electroweak symmetry breaking. Further, theoretical predictions at higher order involving top quarks depend on them or further SM parameters, allowing for their precise extraction from measurements of the production cross section or kinematic observables. Recent measurements by the ATLAS~\cite{bib:ATLAS} and CMS~\cite{bib:CMS} Collaborations reach compatible or higher precision than corresponding theoretical predictions and are presented here in these proceedings.

\section{Top quark mass measurements}
The value of the top quark mass can be measured in multiple ways, depending on its definition and the experimental procedure used. They can be classified into two distinct major categories. For example, the mass can be measured by reconstructing the energy of top quark decay products or extracting it from measured distributions of kinematic observables at detector level by comparison to predictions of multi-purpose Monte-Carlo (MC) generators. Measurements of this type are commonly classified as direct measurements, and the mass value measured is often referred to as MC top quark mass (\mtmc).
They lack a clear theoretical interpretation compared to indirect measurements, where the top quark mass is extracted in a well defined renormalization scheme such as the pole or modified minimal subtraction ($\mathrm{\overline{MS}}$) scheme by comparing sensitive observables to theoretical calculations at fixed order. 
Usually, \mtmc\ is related to be close or equal to the mass as defined in the pole mass scheme, with an additional uncertainty of the order of $1\,\mathrm{GeV}$ coming from the usage of probabilistic MC generators~\cite{bib:massTheoHoang,bib:massTheoNason}.

\begin{figure}[h]
\centering
\includegraphics[width=0.5\textwidth]{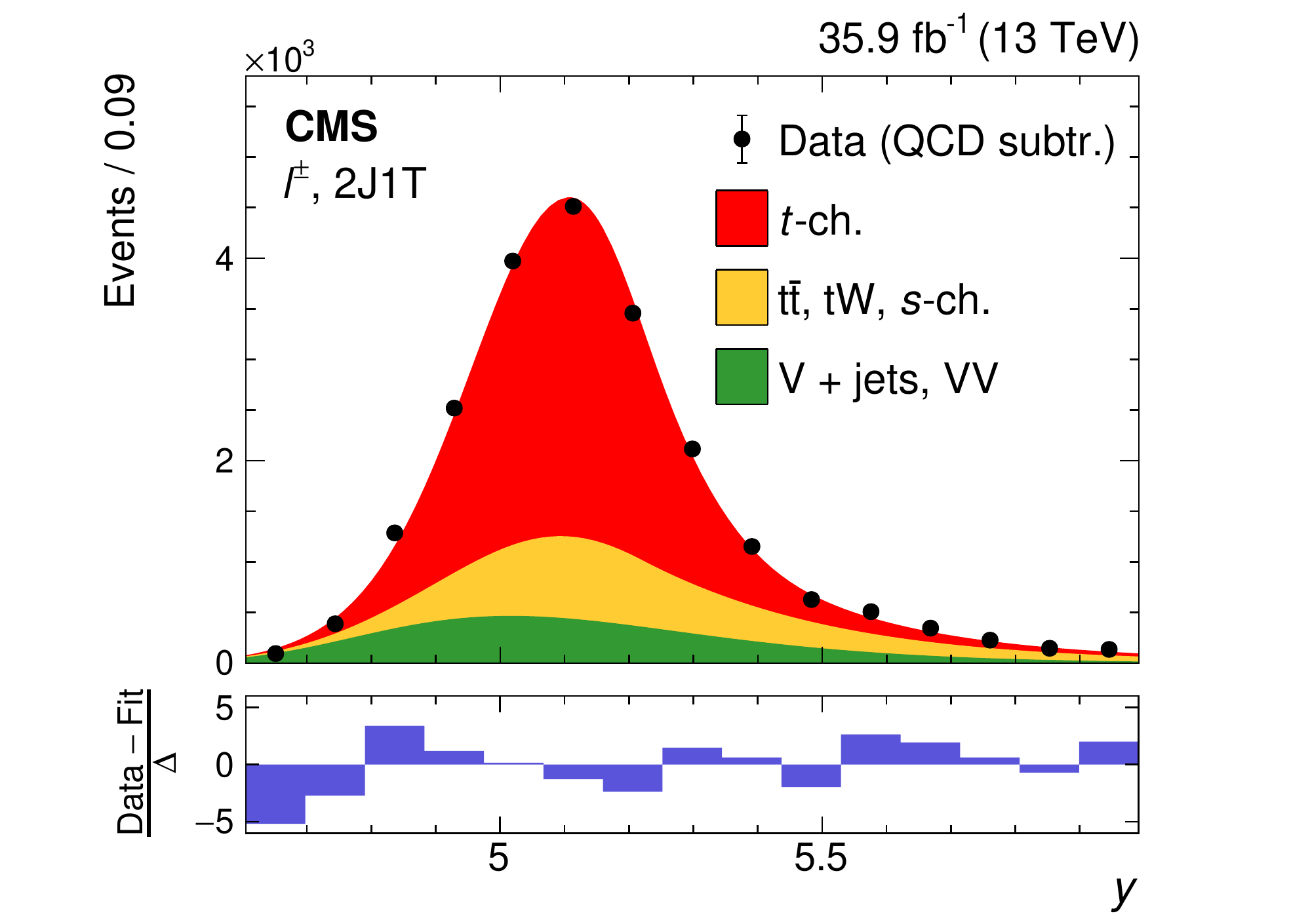}
\includegraphics[width=0.4\textwidth]{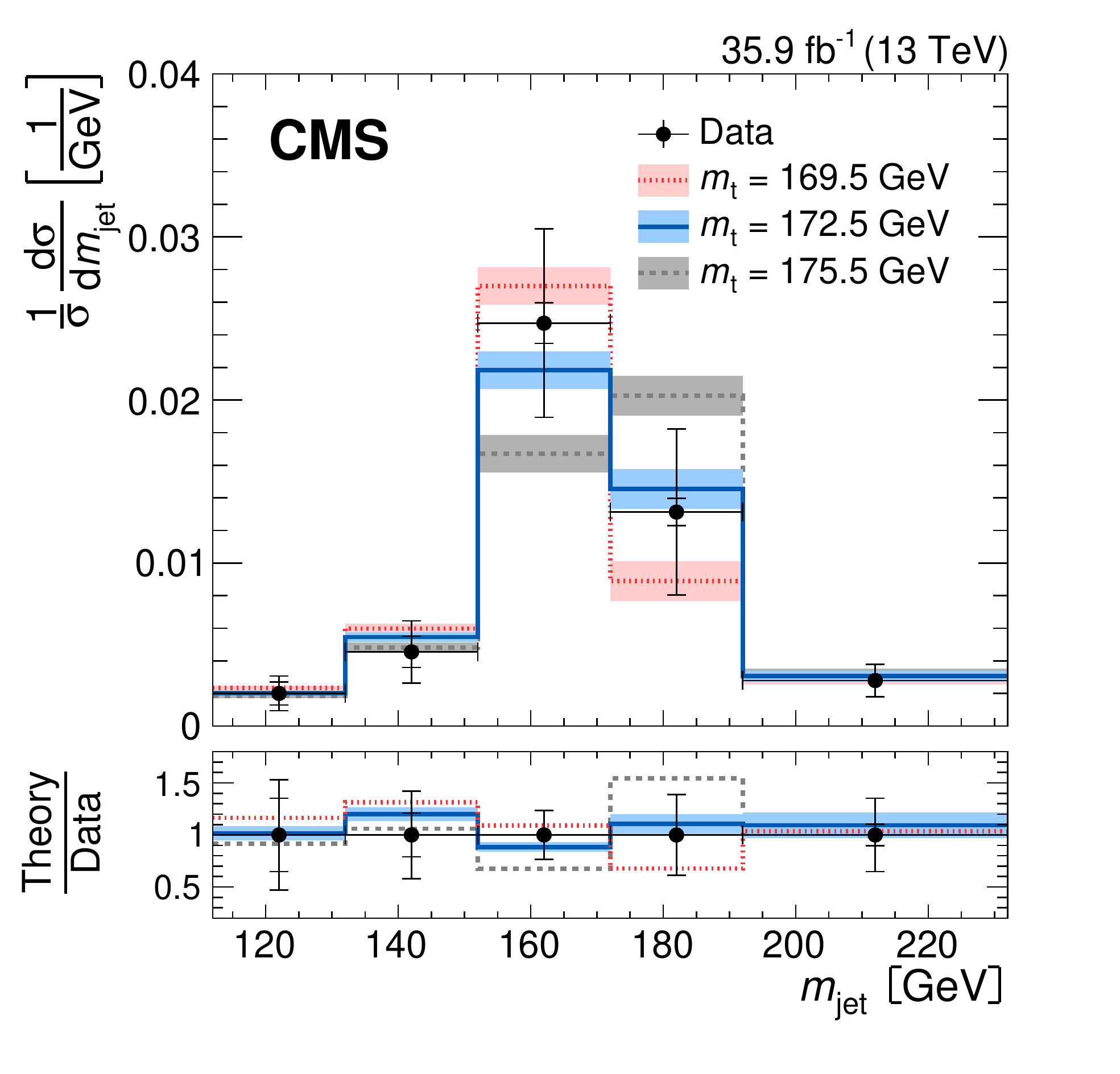}
\caption{Reconstructed data as a function of the natural logarithm of the reconstructed top quark mass together with a functional fit (left)~\cite{bib:singletop}. Unfolded data compared to MC predictions using different top quark mass assumptions as a function of the boosted jet mass. (right)~\cite{bib:boosted}.}
\label{fig:mass1}
\end{figure}

The first sub-GeV precision measurement of the top quark mass in a single top phase space was achieved recently by the CMS Collaboration analyzing single top \textit{t}-channel events using $35.9$~\fbinv\ of pp collision data~\cite{bib:singletop}. Events have been categorized by the charge and flavor of the selected lepton of the subsequent top quark decay and a simultaneous parametric fit was performed to determine the top quark mass. Additionally, a multivariate discriminator was used to enhance the purity of the selected signal events, and two additional fits were performed to determine also the quark and antiquark masses separately. The result of the simultaneous fit is shown in Figure~\ref{fig:mass1} left, and the best fit value for the top quark/antiquark mass yields $172.13 \pm 0.7\, \mathrm{GeV}$. The extracted mass ratio between the quark and antiquark is found to be $0.995\pm0.006$, and their difference is $0.83_{-1.01}^{+0.77}\, \mathrm{GeV}$. Both values are in agreement with the SM expectation.

By using the mass of jets in the boosted regime in the \ttbar\ topology~\cite{bib:boosted}, connections between direct and indirect measurements can be probed~\cite{bib:boostedHoang}. Such a measurement was performed by the CMS Collaboration using the same amount of pp collision data as in the previous analysis. The normalized differential cross section as a function of the boosted jet mass is measured and unfolded to the particle level. By comparing the unfolded data to theoretical predictions, the value for \mtmc\ is extracted to be $172.6 \pm 1.9\,\mathrm{GeV}$ (see Figure~\ref{fig:mass1} right). With respect to previous iterations of the same measurement, the resolution of the measured jet mass distribution and thus the precision of the extraction is significantly improved by the use of the XCone jet clustering algorithm for the first time at the LHC.

\begin{figure}[h]
\centering
\includegraphics[width=0.4\textwidth]{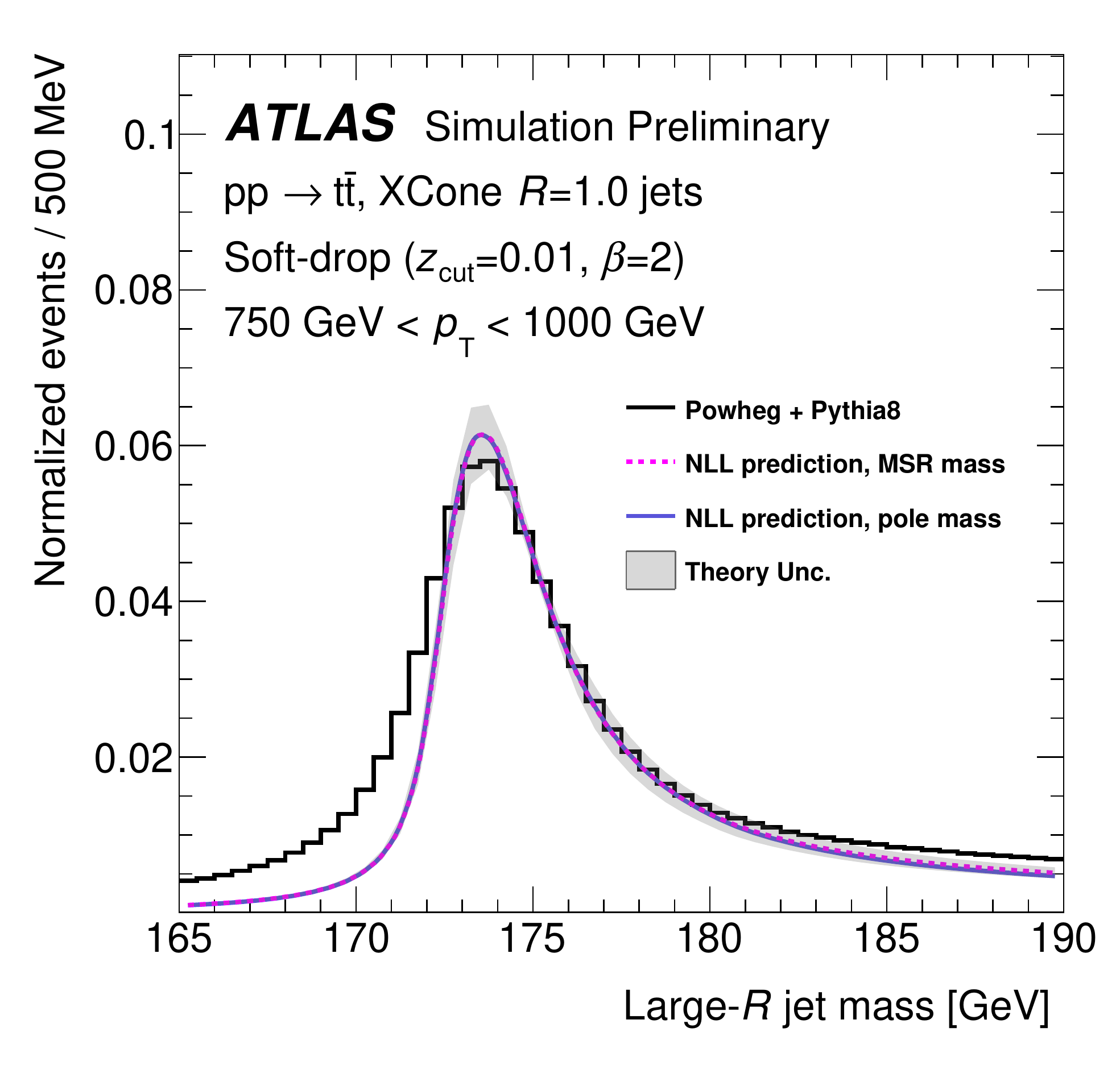}
\includegraphics[width=0.50\textwidth]{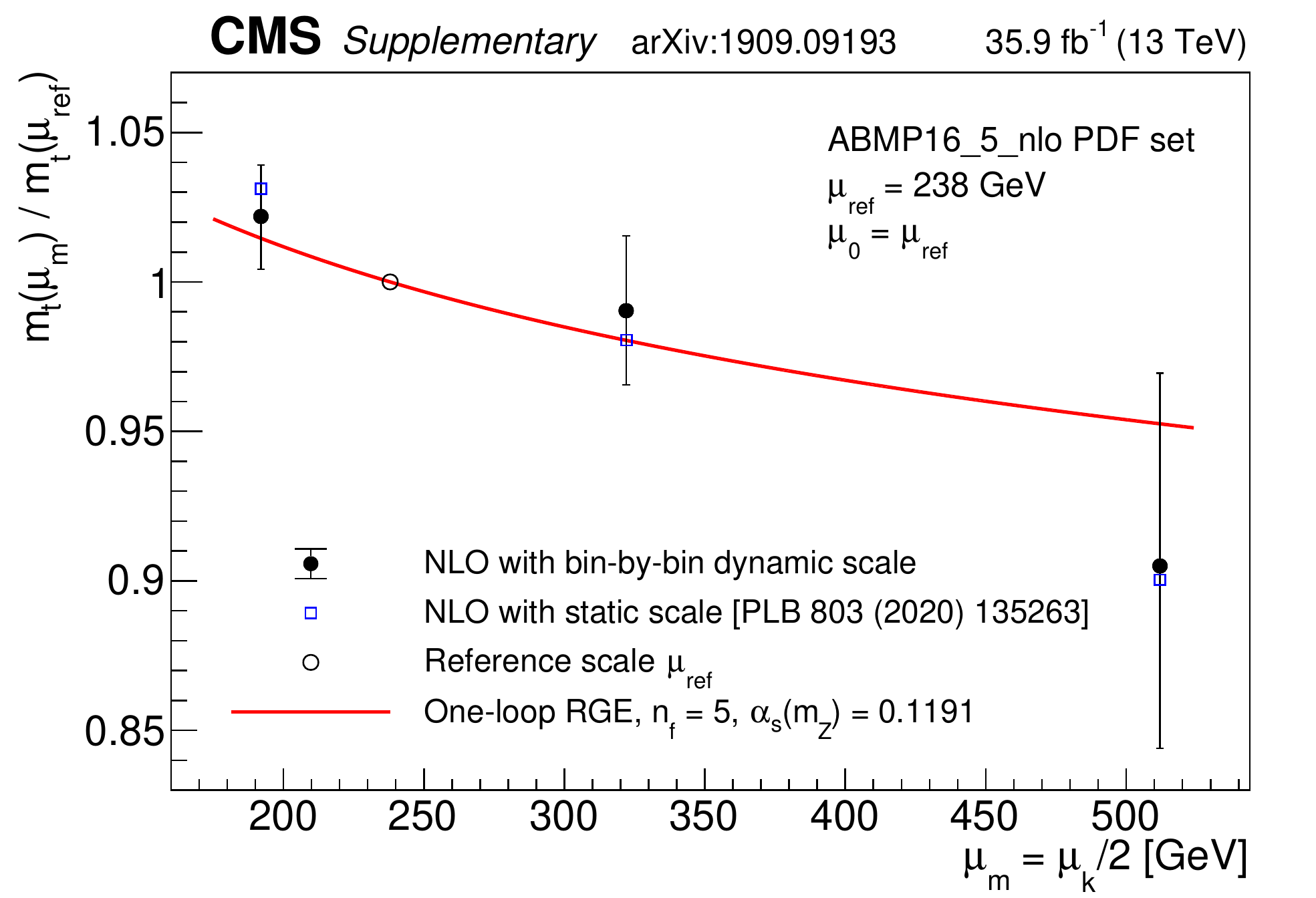}
\caption{Particle-level boosted jet mass distribution compared to best fit results for a NLL theory prediction using two top quark mass schemes (left)~\cite{bib:AtlasBoosted}. Measured scale dependence of the top quark mass as a function of the scale $\mathrm{\mu_{m}}=\mathrm{m_{t\bar{t}}}/2$ (right)~\cite{bib:running}.}
\label{fig:mass2}
\end{figure}

Theoretical calculations at next-to-leading-log (NLL) precision for the same observable are used in a recent interpretation of the ATLAS Collaboration to understand the relation between the ATLAS \mtmc\ parameter and well defined theoretical masses~\cite{bib:AtlasBoosted}. By performing template fits of the NLL theoretical prediction with the top quark mass in the MSR~\cite{bib:MSR} and the pole mass scheme as free parameters to the ATLAS MC simulation at particle level, the numerical difference between \mtmc\ and the two theoretical mass parameters can be extracted. An exemplary result of template fits for the MSR and pole mass is shown in Figure~\ref{fig:mass2} left. The difference between \mtmc\ and $m_{\mathrm{t}}^{\mathrm{MSR}}(R=1\,\mathrm{GeV})$ is found to be $80^{+350}_{-410}\,\mathrm{MeV}$, and $350^{+300}_{-360}\,\mathrm{MeV}$ for the difference between \mtmc\ and $m_{\mathrm{t}}^{\mathrm{pole}}$. A scale of $R=1\,\mathrm{GeV}$ is chosen because of the numerical similarity to the pole mass ($m_{\mathrm{t}}^{\mathrm{MSR}}(R=1\,\mathrm{GeV}) \approx m_{\mathrm{t}}^{\mathrm{pole}}$). 

CMS investigated for the first time the scale dependence (running) of the top quark mass in the $\mathrm{\overline{MS}}$ scheme using \ttbar\ events reconstructed in the dileptonic final state using $35.9$~\fbinv\ of collision data. By performing a profiled maximum likelihood fit to final state observables, the measured invariant mass spectrum of the reconstructed \ttbar\ pair (\mttbar) is unfolded to the parton level. The running is probed up to a scale of 1~TeV by comparing the measured differential cross section to theoretical predictions and is found to be in agreement with the SM within $1.1\sigma$ (see Figure~\ref{fig:mass2} right). 

The top quark mass was further measured by ATLAS analyzing events in which the \ttbar\ pair decays semileptonically~\cite{bib:ATLASmasslep}. Here, the sensitivity of the shape of the invariant mass distribution of a lepton and a soft muon coming from a B~hadron decay to the top quark mass is used. This particular analysis mitigates the dependence on typical uncertainties of top quark mass measurements coming from jet energy calibrations. The top quark mass is found to be $\mt = 174.48 \pm 0.78\,\mathrm{GeV}$.

\section{Probing heavy quark fragmentation}
Two measurements of differential distributions being sensitive to the modeling of bottom quark fragmentation function have been carried out by ATLAS~\cite{bib:ATLASbfrag} and CMS~\cite{bib:CMSbfrag}. Bottom quarks play a vital role in many LHC measurements like in the top quark sector, and measuring the bottom quark fragmentation function is a fundamental test of perturbative QCD and the parton shower formalism. In both measurements, observables are defined based on charged-particle tracks that characterize the bottom-quark momentum relative to the momentum of the jet. In the ATLAS analysis, the measured distributions are corrected for detector effects via unfolding and are compared to predictions obtained from different MC generators. In the CMS measurement, the shape of the distributions is used to extract th best fit value for the Bowler-Lund bottom-quark fragmentation shape parameter $r_b$ for the first time at the LHC. It is found to be $0.858 \pm 0.037 (\mathrm{stat}) \pm 0.031 (\mathrm{syst})$, which is in good agreement with measurements using $\mathrm{e}^{+}\mathrm{e}^{-}$ data performed at the CERN LEP.

\section{Measurements of electroweak parameters}
Electroweak contributions to \ttbar\ production, like the exchange of massive bosons between the final state top quarks, allow to study the dependence of the differential production cross section on fundamental SM parameters such as the top quark Yukawa coupling \yukawa. CMS has performed a measurement of \yukawa\ by performing a profiled likelihood fit to final state distributions using $137\,\fbinv$ of collision data~\cite{bib:yukawa}. To mitigate the dependence of the missing transverse momentum resolution in the dileptonic decay channel, proxy variables are determined. The dependence of the shape of the distributions on different \yukawa\ assumptions are modeled using HATHOR~\cite{bib:hathor} predictions. The best fit value for \yukawa\ is found to be $1.16^{+0.24}_{-0.35}$ and similarly an upper limit of $\yukawa<1.5$ is calculated at 95\%~CL.

By exploiting the dependence of the single top \textit{t}-channel production cross section on the entries of the Cabibbo-Kobayashi-Maskawa (CKM) matrix, their elements $\mathrm{V_{tb}}$, $\mathrm{V_{td}}$, and $\mathrm{V_{ts}}$ can be extracted. This is done in the reported CMS analysis in a model independent and simultaneous way~\cite{bib:CKM}. Events of pp collision data corresponding to an integrated luminosity of $35.9$~\fbinv are separated using multivariate discriminators, and the individual signal strength parameters of different \textit{t}-channel production modes are fitted simultaneously in a profiled maximum likelihood fit. For the interpretation, three different beyond the SM physics scenarios are assumed, and also the top quark decay width is constrained. 

Finally, the reported measurement performed by ATLAS puts a fundamental assumption in the SM to test, probing the ratio of couplings strengths of different fermion generations to gauge bosons~\cite{bib:ATLASlepton}.
Using 139~\fbinv  of data recorded with the ATLAS detector in pp collisions, the ratio of $ R(\tau/\mu)=B(W\to\tau\nu_{\tau})/B(W\to\mu\nu_{\mu})$ is measured the first time at the LHC. Novel methods are used for this kind of measurement, and a value of $0.992\pm0.013[\pm0.007(\mathrm{stat})\pm0.011(\mathrm{syst})]$ is extracted through a template likelihood fit to the spectra of the impact parameter and transverse momentum of the muon. The measured value is found to be in agreement with the SM assumption of lepton universality and surpasses the precision of LEP for the same ratio by a factor of 2.

\section{Summary}
Using measurements of top quark production, fundamental parameters of the standard model QCD Lagrangian, like the top quark mass, or the electroweak sector, e.g., CKM matrix elements and the Yukawa coupling, can be measured. Similarly, heavy-quark fragmentation functions can be probed, and a better understanding of the parton shower formalism and Monte-Carlo multi-purpose generators can be gained. Significant progress was made by the ATLAS and CMS Collaborations using the data collected during LHC Run II at a center of mass energy of $13\,\mathrm{TeV}$ and recent measurements were presented.

%An introduction that also uses index \index{Mesmer}
%And a citation~\cite{Mesmer}. A second citation~\cite{diCenzo}.
%A Figure is shown in Figure~\ref{fig:magnet}. 

%%%%%%%%%%%%%%%%%%%%%%%%%%%%%%%%%%%%%%%%%%%%%%%%%%%%%%%%%%%%%%%%%%%%%%%%%
%%
%%   use this format to include an .eps figure into your paper
%%
%\begin{figure}[!h!tbp]
%\centering
%\includegraphics[height=1.5in]{magnet}
%\caption{Brief figure caption.}
%\label{fig:magnet}
%\end{figure}
%%%%%%%%%%%%%%%%%%%%%%%%%%%%%%%%%%%%%%%%%%%%%%%%%%%%%%%%%%%%%%%%%%%%%%%%%%%

%Some sample blood cyanide levels are reported in Table~\ref{tab:ex}.

%%%%%%%%%%%%%%%%%%%%%%%%%%%%%%%%%%%%%%%%%%%%%%%%%%%%%%%%%%%%%%%%%%%%%%%%%
%%
%%   use this format to include a LaTeX table  into your paper
%%
%\begin{table}[!h!tbp]
%\begin{center}
%\begin{tabular}{l|ccc}  
%col 1 &  col 2 &  col 3 [GeV]&  
%col 4\\ \hline
% row 1  &   0.12     &     10      &     0.1  \\
% row 2 &   0.15     &     100     &  $\pm 10$ \\ \hline
%\end{tabular}
%\caption{Brief table caption.}
%\label{tab:ex}
%\end{center}
%\end{table}
%%%%%%%%%%%%%%%%%%%%%%%%%%%%%%%%%%%%%%%%%%%%%%%%%%%%%%%%%%%%%%%%%%%%%%%%%%%

%\Acknowledgements
%I am grateful.

\bibliography{eprint}{}
\bibliographystyle{unsrt}
 
\end{document}

%% file: econfmacros.tex
\def\beq{\begin{equation}}
\def\eeq#1{\label{#1}\end{equation}}
\def\eeqn{\end{equation}}

\def\beqa{\begin{eqnarray}}
\def\eeqa#1{\label{#1}\end{eqnarray}}
\def\eeqan{\end{eqnarray}}

\let\bar=\overbar

\def\Dslash{\not{\hbox{\kern-4pt $D$}}}
\def\dslash{\not{\hbox{\kern-2pt $\del$}}}

\def\mt{m_t}

\def\msb{{\bar{\ssstyle M \kern -1pt S}}}

%% file: eprint.bbl
\begin{thebibliography}{10}

\bibitem{bib:ATLAS}
{ATLAS Collaboration}.
\newblock The {ATLAS} experiment at the {CERN} {L}arge {H}adron {C}ollider.
\newblock {\em JINST}, 3:S08003, 2008.

\bibitem{bib:CMS}
{CMS Collaboration}.
\newblock {The CMS Experiment at the CERN LHC}.
\newblock {\em JINST}, 3:S08004, 2008.

\bibitem{bib:massTheoHoang}
Andr\'e~H. Hoang, Simon Pl\"atzer, and Daniel Samitz.
\newblock {On the cutoff dependence of the quark mass parameter in angular
  ordered parton showers}.
\newblock {\em JHEP}, 10:200, 2018.

\bibitem{bib:massTheoNason}
Silvia Ferrario~Ravasio, Tom\'a\v{s} Je\v{z}o, Paolo Nason, and Carlo Oleari.
\newblock {A theoretical study of top-mass measurements at the LHC using NLO+PS
  generators of increasing accuracy}.
\newblock {\em Eur. Phys. J. C}, 78(6):458, 2018.
\newblock [Addendum: Eur.Phys.J.C 79, 859 (2019)].

\bibitem{bib:singletop}
Armen Tumasyan et~al.
\newblock {Measurement of the top quark mass using events with a single
  reconstructed top quark in pp collisions at $\sqrt{s}$ = 13 TeV}.
\newblock {\em submitted to JHEP}, 2021.

\bibitem{bib:boosted}
{CMS Collaboration}.
\newblock Measurement of the jet mass distribution and top quark mass in
  hadronic decays of boosted top quarks in pp collisions at
  $\sqrt{s}=13\,\mathrm{TeV}$.
\newblock {\em Phys. Rev. Lett.}, 124:202001, 2020.

\bibitem{bib:boostedHoang}
Sean Fleming, Andre~H. Hoang, Sonny Mantry, and Iain~W. Stewart.
\newblock Jets from massive unstable particles: Top-mass determination.
\newblock {\em Phys. Rev. D}, 77:074010, 2008.

\bibitem{bib:AtlasBoosted}
{ATLAS Collaboration}.
\newblock {A precise interpretation for the top quark mass parameter in ATLAS
  Monte Carlo simulation}, 2021.
\newblock ATL-PHYS-PUB-2021-034 \url{https://cds.cern.ch/record/2777332/}.

\bibitem{bib:MSR}
Andr\'e~H. Hoang, Ambar Jain, Ignazio Scimemi, and Iain~W. Stewart.
\newblock Infrared renormalization-group flow for heavy-quark masses.
\newblock {\em Phys. Rev. Lett.}, 101:151602, 2008.

\bibitem{bib:running}
{CMS Collaboration}.
\newblock Running of the top quark mass from proton-proton collisions at
  $\sqrt{s}=13\, \mathrm{TeV}$.
\newblock {\em Physics Letters B}, 803:135263, 2020.

\bibitem{bib:ATLASmasslep}
{ATLAS Collaboration}.
\newblock {Measurement of the top quark mass using a leptonic invariant mass in
  pp collisions at $\sqrt{s}$ = 13 TeV with the ATLAS detector}, 2019.
\newblock ATLAS-CONF-2019-046 \url{https://cds.cern.ch/record/2693954}.

\bibitem{bib:ATLASbfrag}
{ATLAS Collaboration}.
\newblock {Measurements of b--jet moments sensitive to b--quark fragmentation
  in $\mathrm{t \bar{t}}$ events at the LHC with the ATLAS detector}, 2020.
\newblock ATLAS-CONF-2020-050 \url{https://cds.cern.ch/record/2730444}.

\bibitem{bib:CMSbfrag}
{CMS Collaboration}.
\newblock {Measurement of the shape of the b quark fragmentation function using
  charmed mesons produced inside b jets from $\mathrm{t}\bar{\mathrm{t}}$ pair
  decays}, 2021.
\newblock CMS-PAS-TOP-18-012 \url{https://cds.cern.ch/record/2771694}.

\bibitem{bib:yukawa}
{CMS Collaboration}.
\newblock Measurement of the top quark {Yukawa} coupling from
  $\mathrm{t\overline{t}}$ kinematic distributions in the dilepton final state
  in proton-proton collisions at $\sqrt{s}=13\,\,\mathrm{TeV}$.
\newblock {\em Phys. Rev. D}, 102:092013, 2020.

\bibitem{bib:hathor}
M.~Aliev, H.~Lacker, U.~Langenfeld, S.~Moch, P.~Uwer, and M.~Wiedermann.
\newblock {HATHOR: HAdronic Top and Heavy quarks crOss section calculatoR}.
\newblock {\em Comput. Phys. Commun.}, 182:1034--1046, 2011.

\bibitem{bib:CKM}
{CMS Collaboration}.
\newblock Measurement of {CKM} matrix elements in single top quark t-channel
  production in proton-proton collisions at $\sqrt{s}=13\, \mathrm{TeV}$.
\newblock {\em Physics Letters B}, 808:135609, 2020.

\bibitem{bib:ATLASlepton}
{ATLAS Collaboration}.
\newblock {Test of the universality of $\tau$ and $\mu$ lepton couplings in
  $W$-boson decays with the ATLAS detector}.
\newblock {\em Nature Phys.}, 17(7):813--818, 2021.

\end{thebibliography}
